%% file: main.tex
\documentclass[runningheads]{llncs}

\usepackage{graphicx} 
\usepackage{bmpsize}

\usepackage{xcolor}

\usepackage{amssymb}
\usepackage{multicol}
\usepackage{multirow}
\usepackage{bm}

\usepackage{url}

\usepackage{subfigure}

\begin{document}

\title{Synthetic is all you need: removing the auxiliary data assumption for membership inference attacks against synthetic data}
\titlerunning{Synthetic is all you need}

\author{Florent Guépin \textsuperscript{*} \and
Matthieu Meeus \textsuperscript{*} \and
Ana-Maria Cre\c{t}u \and Yves-Alexandre de Montjoye  
}

\authorrunning{F. Guépin, M. Meeus et al.}

\institute{Department of Computing and Data Science Institute, 
Imperial College London, London, United Kingdom \\
\email{\{florent.guepin20,m.meeus22,a.cretu,deMontjoye\}@imperial.ac.uk}}

\maketitle              

\def\thefootnote{*}\footnotetext[1]{These authors contributed equally to this work}

\begin{abstract}
Synthetic data is emerging as one of the most promising solutions to share individual-level data while safeguarding privacy. While membership inference attacks (MIAs), based on shadow modeling, have become the standard to evaluate the privacy of synthetic data, they currently assume the attacker to have access to an auxiliary dataset sampled from a similar distribution as the training dataset. This is often seen as a very strong assumption in practice, especially as the proposed main use cases for synthetic tabular data (e.g. medical data, financial transactions) are very specific and don't have any  reference datasets directly available. 
We here show how this assumption can be removed, allowing for MIAs to be performed using only the synthetic data. Specifically, we developed three different scenarios: (S1) Black-box access to the generator, (S2) only access to the released synthetic dataset and (S3) a theoretical setup as upper bound for the attack performance using only synthetic data. Our results show that MIAs are still successful, across two real-world datasets and two synthetic data generators. These results show how the strong hypothesis made when auditing synthetic data releases – access to an auxiliary dataset – can be relaxed, making the attacks more realistic in practice.

\end{abstract}

\section{Introduction}
\label{sec:introduction}
\input{sections/introduction}

\section{Background and Related Work}
\label{sec:background}

\input{sections/background}

\section{Attack scenarios}
\label{sec:methods}
\input{sections/methods}

\section{Experimental Setup}
\label{sec:experimental_setup}
\input{sections/experimental_setup}

\section{Results}
\label{sec:results}
\input{sections/results}

\section{Future Work}
\label{sec:future_work}
\input{sections/future_work}

\section{Conclusion}
\label{sec:conclusion}
\input{sections/conclusion}

\subsubsection{Acknowledgements} 

We acknowledge computational resources and support provided by the Imperial College Research Computing Service\footnote{\url{http://doi.org/10.14469/hpc/2232}.}.

\bibliographystyle{splncs04}
\bibliography{mybibliography}
\end{document}

%% file: sections/introduction.tex
Data is crucial in statistical modeling, machine learning systems, and decision-making processes, driving research and innovation. However, data often pertains directly or indirectly to individuals and may contain sensitive information, such as medical records and financial transactions, raising privacy concerns.

Synthetic tabular data is a promising solution to share data while limiting the risk of re-identification \cite{bellovin2019privacy}.
A synthetic data generator is a statistical model trained on the original, private dataset and used to generate synthetic records. The generated synthetic records would then not be linkable to any specific individual while retaining most of the statistical utility of the original dataset.
Extensive research has been dedicated to exploring a wide range of techniques for generating synthetic data \cite{zhang2017,synthpop2022,ctgan2019,jordon2019pate,reprosyn2022}. 
Since, if truly anonymous, synthetic data would fall outside the scope of data protection legislation such as the European Union's General Data Protection Regulation (EU GDPR) \cite{eu-gdpr} or California Consumer Privacy Act \cite{ccpa}, various sectors including finance \cite{fca22synthetic}, healthcare \cite{tucker2020generating}, and research \cite{edge2020design} have expressed significant interest in its adoption in practice. 

However, synthetic data alone does not necessarily preserve privacy. 
First, it is long known that aggregation alone does not effectively safeguard privacy \cite{dinur2003revealing,pyrgelis2017knock}. Second, achieving formal privacy guarantees for synthetic data generation models poses implementation challenges and currently comes at a cost in utility \cite{stadler2022synthetic,annamalai2023linear}. 

Membership inference attacks (MIAs) have thus been used to evaluate the privacy preservation capabilities of synthetic data in practice.
An MIA aims to infer if a specific target record is part of the generative model's training set. Recent work has shown that synthetic data is vulnerable to MIAs, with state-of-the-art attacks relying on the shadow modeling approach \cite{shokri2017membership,stadler2022synthetic,houssiau2022tapas}. 
This approach involves training a membership classifier to distinguish between synthetic datasets generated from so-called shadow datasets with or without a particular target record. 
Importantly, these attacks require the attacker to have access to an auxiliary dataset that follows the same distribution as the original, private dataset, from which the attacker will sample their shadow datasets.

We here argue that this is often a strong assumption in practice \cite{salem2018ml}.
While general datasets of images are widely available, medical datasets or datasets of financial transactions -- some of the main use cases for synthetic tabular data -- are not only not widely available but also very specific e.g. to certain geographies, type of diseases, etc.
The practical feasibility of an attack is also an important criterion from a legal perspective when assessing what constitutes anonymous data.
Recital 26 of the EU GDPR~\cite{eu-gdpr} indeed states that ``account should be taken of \textit{all the means reasonably likely to be used}, such as singling out, either by the controller or by another person to identify the natural person directly or indirectly.''

\textbf{Contribution}.
In this work, we show how synthetic data can effectively replace the auxiliary dataset when running MIAs, removing the strong assumption made by attacks so far and making our attack --in our opinion-- more reasonably likely from a legal perspective.

First, we consider an attacker with black-box access to the synthetic data generator, which is used to generate shadow datasets for running the MIA.
We evaluate the shadow modeling attacks of Houssiau et al.~\cite{houssiau2022tapas} and Meeus et al.~\cite{meeus2023achilles} on two real-world datasets, two synthetic data generators and across ten target records identified by the vulnerable record identification method of Meeus et al. \cite{meeus2023achilles}.
Our results show that MIAs based on synthetic data alone leak the membership of their most vulnerable records 65.5\% of the time on average across datasets and generators.
This is 15.5 percentage points (p.p.) better than the random guess baseline.
We then compare the MIA performance to the traditional setting that assumes access to an auxiliary dataset from the same distribution.
We find that our attacker only loses 11.6 p.p. when compared to this much stronger assumption.

Second, we consider an even weaker attacker that exclusively uses the released synthetic data to perform shadow modeling-based MIAs. 
This attacker obtains an average accuracy of 62.8\%. This result is especially meaningful as having access to the released synthetic dataset is an assumption almost always met in practice. Even here, we show the attack to still work 12.8 p.p. better than the random guess baseline. 

Third, we identify a potential \emph{double counting} issue which might lower the accuracy of an attack when using synthetic data to replace the auxiliary dataset. We formalize the problem and propose an empirical setup, where we artificially solve the double counting issue, to compute an upper-bound on the accuracy of an attack using only synthetic data. We show the upper-bound to reach 85.8\%, 8.7 p.p. higher than the auxiliary data scenario, emphasizing how synthetic only attacks might in the future outperform what is today considered the risk posed by a strong attacker.

MIAs are the main tool to evaluate the privacy-preserving capabilities of synthetic data. However, the strong auxiliary data assumption they currently rely on might lead some to question the practical risk posed by these attacks \cite{salem2018ml,deng2022understanding} and whether they are 'reasonably likely'. We here show how this assumption can be relaxed, as attackers having solely access to the synthetic data generator or even released synthetic data are still able to develop meaningful attacks. We finally find that future attacks based on synthetic data might outperform traditional attacks if the double counting issue can be resolved.

%% file: sections/background.tex
\subsection{Synthetic data generation}
\label{subsec:background:synthetic_data}
Suppose that an entity (e.g. governmental institution, company) seeks to grant a third party access to a private, tabular dataset $D$ for analysis. This dataset consists of a collection of records, each corresponding to an individual, which we denote by $D = \{x_1, \ldots, x_n\}$. 
Each record consists of $F$ features, where a feature is the value for a given attribute.

To address privacy risks, realizing that anonymizing record-level data often fails~\cite{de2013unique}, an increasingly popular approach involves training a synthetic data generator and publishing a synthetic dataset \cite{bellovin2019privacy}. Synthetic data is created by (1) fitting a statistical model to the original data, and (2) using this model to generate artificial (``synthetic'') records by sampling new values.
Ideally, the synthetic data should preserve key statistical properties of the original dataset $D$ without disclosing private information of the individuals in $D$.

The statistical model employed for generating synthetic data is referred to as the \textit{synthetic data generator} $\phi$, and we write $D^s \sim \Phi, |D^s|=m$ to denote that a synthetic dataset of $m$ records is sampled i.i.d. from the generator $\Phi$, fitted on a dataset $D$.
We write $\Phi = \mathcal{G}(D)$ to say that a certain fitting procedure $\mathcal{G}$ (e.g., parameter fitting of a Bayesian network) was applied to the original dataset $D$ to obtain the generator $\Phi$.
The generator can take various forms, such as a probabilistic model like Bayesian networks (BayNet) \cite{zhang2017} and Synthpop \cite{nowok2016synthpop} or a generative adversarial network like CTGAN \cite{ctgan2019}.

\subsection{Membership inference attacks against synthetic tabular data}
\label{subsec:background:mia}

Membership inference attacks (MIAs) have become the standard to evaluate the privacy of synthetic data, machine learning (ML) models, and aggregation mechanisms more broadly.
Given the output of an aggregation mechanism, e.g., a synthetic dataset or a set of aggregate statistics computed on a private dataset $D$, the aim of an MIA is to infer whether a given target record $x_T$ was part of $D$ or not. Successful MIAs have been developed against aggregate statistics of e.g. location data \cite{pyrgelis2017knock}, genomic data \cite{homer2008resolving,sankararaman2009genomic}, and against ML models \cite{shokri2017membership,salem2018ml,carlini2022membership}.

For MIAs against synthetic tabular data, a first class of methods directly compares the synthetic records to the original records, searching for exact or near-matches \cite{domingo2015disclosure,yale2019assessing,yale2019privacy,giomi2022unified}.
Stadler et al. \cite{stadler2022synthetic} argue, however, that the studies relying on similarity testing severely underestimate the risk and instead propose an attack using the shadow modeling approach.
First introduced to evaluate the privacy of ML models \cite{shokri2017membership}, the shadow modeling approach is now the state-of-the-art in evaluating the privacy of synthetic data \cite{stadler2022synthetic,houssiau2022tapas,meeus2023achilles}. 

\begin{figure}[!htbp]
    \centering
    \includegraphics[width = 0.5\linewidth]{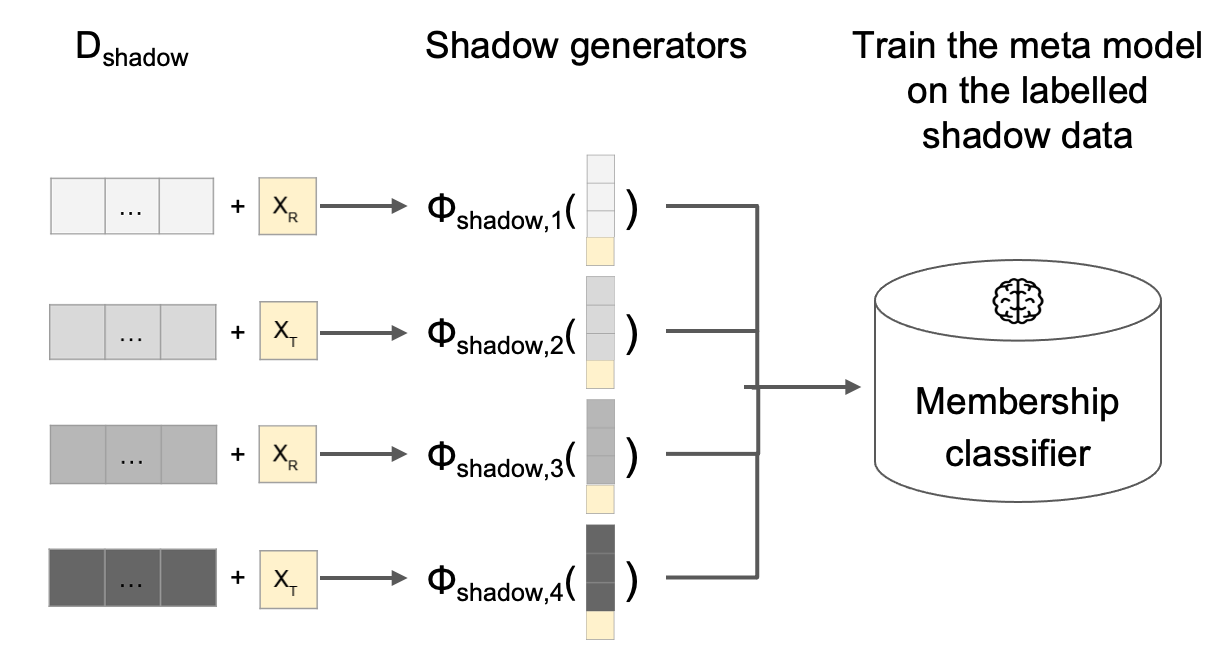}
    \caption{Illustration of the shadow modeling technique}
    \label{fig:shadow_modeling}
\end{figure}

Shadow modeling typically assumes that the attacker has knowledge of the model $\Phi_T$ used to generate the synthetic data and has access to an auxiliary dataset $D_{aux}$ that comes from the same distribution as the original dataset ($D_{aux} \sim \mathcal{D}$). The attacker then constructs multiple shadow datasets $D_{shadow}$ utilizing $D_{aux}$. First, the attacker randomly samples $|D|-1$ records from $D_{aux}$, to then add the target record $x_T$ to 50\% of the shadow datasets, and a random record $x_R$ to the other 50\% instead.
Next, by using the knowledge of the model $\Phi_T$, the attacker will train multiple shadow generators $\Phi_{shadow}$, which in turn produce synthetic shadow datasets $D_{shadow}^s$. 
The attacker knows which $D_{shadow}^s$ have been derived from a shadow dataset containing the target record $x_T$ and which were not. This enables the attacker to train a binary meta-classifier on features extracted from the synthetic shadow datasets to predict membership. Figure \ref{fig:shadow_modeling} illustrates how the shadow modeling technique is used to train the meta-classifier. Lastly, the meta-classifier is evaluated on synthetic test datasets that are similarly constructed (with 50\% having seen the target record during training), but on a disjoint set of records. 

Different techniques have been proposed to extract meaningful features from the synthetic shadow datasets to predict membership. Stadler et al.~\cite{stadler2022synthetic} proposed to extract aggregate statistics, specifically the mean and standard deviation of the attributes, and correlation matrices and histograms.
Houssiau et al.~\cite{houssiau2022tapas} extended this work with a \emph{query-based} feature extractor, using $k$-way marginal statistics computed over the values of the target record for randomly selected subsets of attributes. 
Lastly, Meeus et al. \cite{meeus2023achilles} developed the first trainable feature extractor, which uses the synthetic dataset directly as input to an attention-based classifier. The authors compared the two approaches, showing that the \emph{query-based} method is the state-of-the-art attack on tabular records.  

In previous work, attacks against machine learning models using synthetically generated data have been developed \cite{shokri2017membership,crectu2021correlation}.
In one experiment, Shokri et al.~\cite{shokri2017membership} assumed knowledge of the dataset marginals in order to generate synthetic data.
In another experiment, the same authors generated this data using local search techniques but the method was shown to only be effective when applied to binary records~\cite{salem2018ml}.
Finally, Cretu et al.~\cite{crectu2021correlation} generated synthetic datasets using the copula generative model that satisfy a subset of the correlations present in the private training dataset $D$.
Differently from these approaches targeting ML models, our work concerns attacks targeting synthetic data and makes no additional assumptions on the attacker's knowledge about the original dataset.

%% file: sections/methods.tex
We exclusively consider state-of-the-art MIAs, which are based on the shadow modeling technique.
We assume that the attacker has access to the synthetic dataset $D^s \sim \phi_T(\mathcal{G}(D))$, where $\phi_T$ is referred to as the target generator. We will refer to the size of the synthetic dataset as $n_{synthetic}$.  
The attacker aims to infer whether a particular record, referred to as the target record $x_T$, was part of the original dataset, i.e., whether $x_T\in D$ or $x_T \notin D$.
In line with the literature, we consider the standard setting under which the attacker knows the fitting procedure $\mathcal{G}$ used to train the statistical model on the original data.

To model the uncertainty of the attacker about the dataset, we consider four attack scenarios. First, \textbf{(S0) Auxiliary} is the traditional setup where the attacker has access to an auxiliary dataset sampled from the same distribution as the private dataset. We then propose two new scenarios assuming a weaker attacker: \textbf{(S1) Black-box}, where the attacker has access to the target generator $\phi_T$ and can query the generator an arbitrary number of times to sample synthetic records and \textbf{(S2) Published}, where the attacker has only access to a released synthetic dataset $D^s$ of the same size as the private dataset. Lastly, as a fourth scenario, we construct an artificial setup \textbf{(S3) Upper bound} to evaluate the upper bound of MIAs against synthetic data while only using synthetic records.

\subsection{(S0) Auxiliary}
As a baseline, we consider the traditional attack scenario 
\cite{houssiau2022tapas,stadler2022synthetic} where the attacker has access to an auxiliary dataset $D_{aux}$ sampled from the same distribution $\mathcal{D}$ as the private dataset $D$, i.e. $D_{aux} \sim \mathcal{D}$. $D_{aux}$ is then used to construct the $n_{shadow}$ shadow datasets by uniformly sampling records from $D_{aux}$ without replacement. The meta-classifier is then trained to predict membership with as input features extracted from the synthetic shadow datasets.

Next, the meta-classifier is evaluated on $n_{test}$ synthetic datasets, synthesized from test data that is disjoint from the data used for training. 
The binary membership prediction is then aggregated across all $n_{test}$ synthetic datasets to a final accuracy used as the MIA performance metric. 

\subsection{(S1) Black box}
Next, we remove the auxiliary dataset assumption.
We consider an attacker who is able to query the target generator $\Phi_T$ for synthetic records, i.e. has black-box access to the generator. 
This scenario could, for instance, arise when the end user of the synthetic data would require access to more synthetic records than there were present in the original dataset, e.g. to train ML models. The attacker will here use the black box access to generate $m$ synthetic records to directly construct the shadow datasets. 

Note that, unlike the baseline setting (S0) Auxiliary, the shadow datasets and (consequently) the meta-classifier are now specific to the target generator on which it is evaluated. 
In other words, this setup requires the attacker to train $n_{shadow} \times n_{test}$ generators and $n_{test}$ meta-classifiers while in the standard setting (S0), an attacker needs to train  $n_{shadow} + n_{test}$ generators and only one meta-classifier. 

Again, an attacker will query the trained meta-classifier for one binary prediction for membership per test dataset, which we aggregate to a final accuracy across all $n_{test}$ generators. 

For computational reasons, we sample $m > |D|$ synthetic records for every target generator $\Phi_T$, which we use to sample the shadow datasets in our experiments. 

\subsection{(S2) Published}
In this scenario, we further remove the access to the target generator $\Phi_T$ assumption.
The only knowledge about the original data available to the attacker is the released synthetic dataset $D^s$.
We here assume that the size of the released synthetic dataset is the same as the original, private set, formally $n_{synthetic}=|D^s|=|D|$. 

In this scenario, the attacker trains another generator $\Phi_S$, using the synthetic dataset as training points, i.e., $\Phi_S = \mathcal{G}(D^s)$. With this new generator $\Phi_S$, the attacker generates $m$ new synthetic records to be used to construct the shadow datasets.

We evaluate the MIA performance for this scenario in the same way as in scenario (S1) Black box above. 

\subsection{(S3) Upper bound}

When an MIA against synthetic data for a particular target record is successful, the meta-classifier is able to distinguish whether the target record was part of the original dataset or not. In other words, the meta-classifier is able to detect the effect of the presence of the target record in the original dataset on the generated synthetic data. As shown by Meeus et al. \cite{meeus2023achilles}, this effect is more significant for records more distant to their closest neighbours. 

In scenarios (S1) and (S2), the attacker uses this synthetic data to construct the shadow datasets. When the target record $x_T$ was part of the target generator's training dataset, we hypothesize that using these synthetic records to construct the shadow datasets could deteriorate the performance of the meta-classifier in two ways.
First, as we use synthetic data that is likely impacted by the presence of $x_T$ already to construct the shadow datasets, the two ``worlds'' (presence or absence of $x_T$) in the shadow datasets are likely to be less distinguishable overall by the meta-classifier. Second, this could create a discrepancy in the training (on the shadow datasets) and inference task (on the target generator) of the meta-classifier. We call both effects the \textit{double counting issue} and hypothesize that this could impact the attack performance. 

We formalize this issue by first defining the concept of adjoining synthetic datasets to then define the \textit{trace} of $x_T$.

\begin{definition}
     Let $D = (x_1,\cdots,x_n)$ be a dataset, then an \textbf{adjoining dataset} with respect to $x_T$ will be such that $\exists~k~|~ D^T = (x_1,\cdots, x_k,\bm{x_T},x_{k+2},\cdots,x_n)$ and $x_{k+1} \neq x_T$. We call \textbf{adjoining synthetic datasets} the resulting synthetic datasets generated by the same generator model $\mathcal{G}$ trained on the respective datasets. Namely, $D^{s,T} \sim \Phi = \mathcal{G}(D^T)$ and $D^s \sim \Phi = \mathcal{G}(D)$ are called two adjoining synthetic datasets. 
\end{definition}

\begin{definition}
    Let $\mathcal{D}^s$ and $\mathcal{D}^{s,T}$ be two adjoining synthetic datasets with respect to $x_T$. Then, the \textbf{trace} of $x_T$ is defined as the impact of excluding (respectively including) the target record in the training dataset $D$ ($D\cup \{x_T\} = D^T$) of a synthetic data generator $\Phi = \mathcal{G}(D)$($\Phi = \mathcal{G}(D^T)$) on the generated synthetic data $D^s \sim \Phi$ ($D^{s,T} \sim \Phi$), written $|.|_{\Phi}$. 
    Formally, $\textit{trace}(x_T) = |\mathcal{D}^s - \mathcal{D}^{s,T}|_{\Phi}$. 
\end{definition}

At inference time, the meta-classifier is expected to recognize the trace of $x_T$ i.e. $|\mathcal{D}^{s} - \mathcal{D}^{s,T}|_{\Phi}$. When synthetic data is used to construct the shadow datasets and the target record has not been part of the training data for the target generator, the meta-classifier has been 
 trained to recognize this same trace and hence, the attacker does not encounter the double counting issue. 

However, when the target generator has seen the target record during training, the attacker uses the synthetic dataset $\mathcal{D}^{s,T}$ to construct shadow datasets, each of which will contain $x_T$ with 50\% probability as well. In other words, the synthetic shadow datasets will now be either $\mathcal{D}^{s}_2 \sim \Phi = \mathcal{G}(\mathcal{D}^{s,T} \cup \{x_{random}\})$ or $\mathcal{D}^{s,T}_2 \sim \Phi = \mathcal{G}(\mathcal{D}^{s,T} \cup \{x_T\})$ with 50\% probability. The meta-classifier is hence trained to recognize the trace of trace of $x_T$ i.e. $|\mathcal{D}^{s}_2 - \mathcal{D}^{s,T}_2 |_{\Phi}$, while at inference time it is still expected to recognize the trace of $x_T$, i.e.  $|\mathcal{D}^{s} - \mathcal{D}^{s,T}|_{\Phi}$.

To avoid the double counting issue, we here design a hypothetical attack as a slight modification from scenario (S1). We now artificially ensure that the target $x_T$ is never seen during the training of the generator, to then use the same setup as in (S1). Specifically, when the target is not seen during training, nothing changes, and the attacker has black box access to $\Phi$. In contrast, for a target generator that has seen the target record during training (the target generator will generate $\mathcal{D}^{s,T}$ with $D^T$ as training dataset), we ensure the attacker to have access to an adjoining synthetic dataset $\mathcal{D}^s$, by training the same generator $\Phi$ on an adjoining dataset $D$ of $D^T$.  

This scenario serves as an \textbf{upper bound} for an MIA with access only to synthetic data, since now we artificially avoid the double counting issue. We further evaluate this scenario in the same way as in scenario (S1). 

%% file: sections/experimental_setup.tex
In this section, we describe the experimental setup for the attacks: the synthetic data generation models, datasets, the meta-classifier methods used and the exact attack parameters.

\subsection{Synthetic data generators}

\textbf{Synthpop} has been introduced by Nowok et al. \cite{nowok2016synthpop} as an R package for synthetic data generation. It uses classification and regression trees to estimate conditional probabilities from the training dataset, then used to generate synthetic data. In our work, we utilize the Python re-implementation of Synthpop \cite{synthpop2022} from the reprosyn repository \cite{reprosyn2022}.

\textbf{BayNet} uses a Bayesian Network to generate synthetic data. It represents the attributes of the training data as a Directed Acyclic Graph, capturing causal relationships. Each node in the graph has a conditional distribution $\mathbb{P}[X|Parents(X)]$ estimated from the available data. Synthetic data is generated by sampling from the joint distribution obtained by multiplying the computed conditionals. We also use the implementation from the reprosyn repository \cite{reprosyn2022}.

\subsection{Real world datasets}

\textbf{UK Census}, or the 2011 Census Microdata Teaching File \cite{census2011}, was published by the Office for National Statistics and consists of a random sample representing 1\% of the 2011 Census output database for England and Wales. This dataset comprises a total of $n=569,741$ records and includes $F=17$ categorical attributes. 

\textbf{Adult} \cite{adult1996} is extracted from the 1994 US Census database. The dataset comprises $n=45,222$ records with $F=15$ attributes, 9 of which are categorical and 6 continuous. 

\subsection{Meta-classifier methods}
\label{sec:attacks}
We consider two previously proposed methods to extract features from the synthetic shadow datasets and to train the meta-classifier.

\textbf{Query based}. Introduced by Houssiau et al. \cite{houssiau2022tapas}, this state-of-the-art attack uses $k$-way marginal statistics, or count queries, computed over subsets of the attribute values of the target record from the synthetic dataset. We use $100,000$ randomly sampled count queries of the $2^F$ possibilities and use a random forest classifier with 100 trees and maximum depth of 10 to predict membership.

\textbf{Target Attention}. Introduced by Meeus et al. \cite{meeus2023achilles}, this method takes as input (part of) the synthetic dataset and is the first trainable feature extractor for MIAs against synthetic data. 
The method first computes record-level embeddings. Next, through a custom attention mechanism, these embeddings are aggregated to a dataset-level embedding, which is used to predict binary membership. We use the exact implementation and parameters as laid out in the paper~\cite{meeus2023achilles}.

\begin{table}[htbp!]
    \centering
    \begin{tabular}{|c|c|c|c|c|c|c|}
    \hline
         Dataset & Scenario & \; $|D_{aux}|$ \; & \; $|D_{test}|$ \; & $m$ & $n_{shadow}$ & $n_{test}$ \\
         \hline
        \multirow{4}{*}{Adult} & S0 & 10000 & 5000 & 1000 & \multirow{4}{*}{2000} & \multirow{4}{*}{100} \\
         & S1 & 0 & 5000 & 20000 & & \\
         & S2 & 0 & 5000 & 1000 & & \\
         & S3 & 0 & 5000 & 20000 & & \\
        \hline
        \multirow{4}{*}{UK Census} & S0 & 50000 & 25000 & 1000 & \multirow{4}{*}{2000} & \multirow{4}{*}{100} \\
         & S1 & 0 & 25000 & 20000 & & \\
         & S2 & 0 & 25000 & 1000 & & \\
         & S3 & 0 & 25000 & 20000 & & \\
        \hline
    \end{tabular}
    \caption{Parameters used throughout experiments.}
    \label{tab:parameters}
\end{table}

\subsection{Parameters of the attack}

Table~\ref{tab:parameters} shows the parameters used throughout our experiments. Here, $D_{aux}$ represents the auxiliary dataset and $D_{test}$ the dataset that is used to sample the test datasets.  Both are random and disjoint subsets of the entire dataset. Further, $m$ represents the number of synthetic records queried from the trained generator, $n_{shadow}$ the number of shadow datasets used for training the meta-classifier, and finally $n_{test}$ the number of datasets used for testing.

Throughout our experiments, the size of the released synthetic dataset is equal to the size of the private dataset $D$, i.e., $n_{synthetic} = |D| = 1000$, and similarily for the shadow datasets, i.e. $|D_{shadow}| = |D_{shadow}^s| = 1000$. In scenarios (S1) and (S3) where $m > n_{synthetic}$, we train the meta-classifier using shadow datasets randomly sampled from the $m$ synthetic records. At inference time, we use a random subset of $n_{synthetic} = 1000$ synthetic records to query the trained meta-classifier. 

When constructing both the $n_{shadow}$ shadow datasets for training and $n_{test}$ datasets for testing, we ensure that the target record $x_T$ is present with 50\% probability. This ensures that the evaluation of the attack on the $n_{test}$ datasets is balanced, with a random guess baseline of 50\% accuracy for binary prediction of membership. 

Lastly, for each dataset, we run the attack on $10$ target records selected using the vulnerable record identification method proposed by Meeus et al. \cite{meeus2023achilles}. For each record in the original dataset, the method computes its vulnerability score as the mean cosine distance, generalized across attribute types, to its five closest neighbours. The records that are the most distant from their closest neighbours, i.e. have the largest mean distance, are selected as vulnerable records.

%% file: sections/results.tex
In this section, we evaluate how the performance of the MIA varies across our attack scenarios over two synthetic data generators and two datasets.

\begin{figure}[!ht]
\centering

\subfigure[UK Census]{
\includegraphics[width=0.45\linewidth]{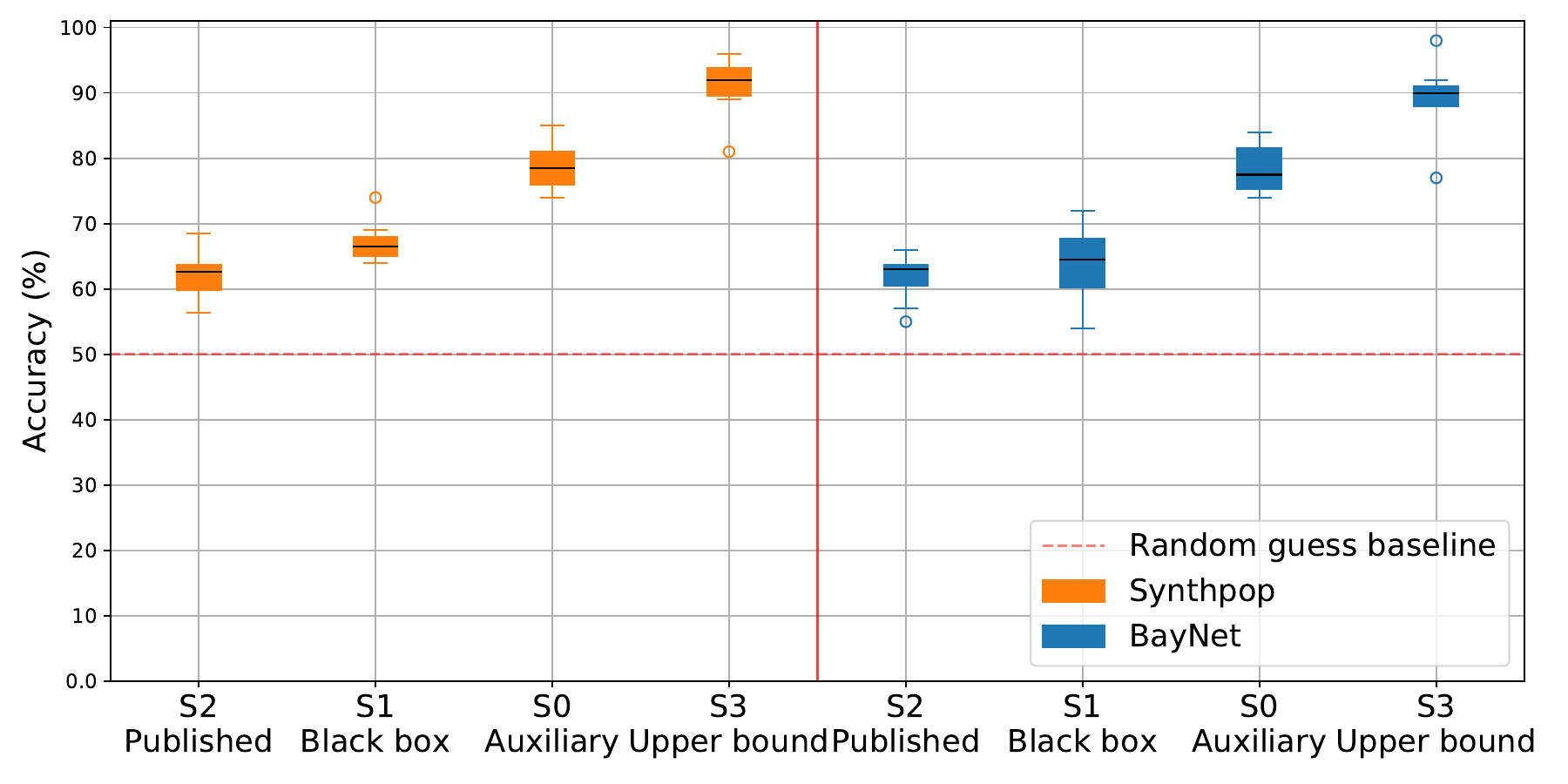}
}
\subfigure[Adult]{
\includegraphics[width=0.45\linewidth]{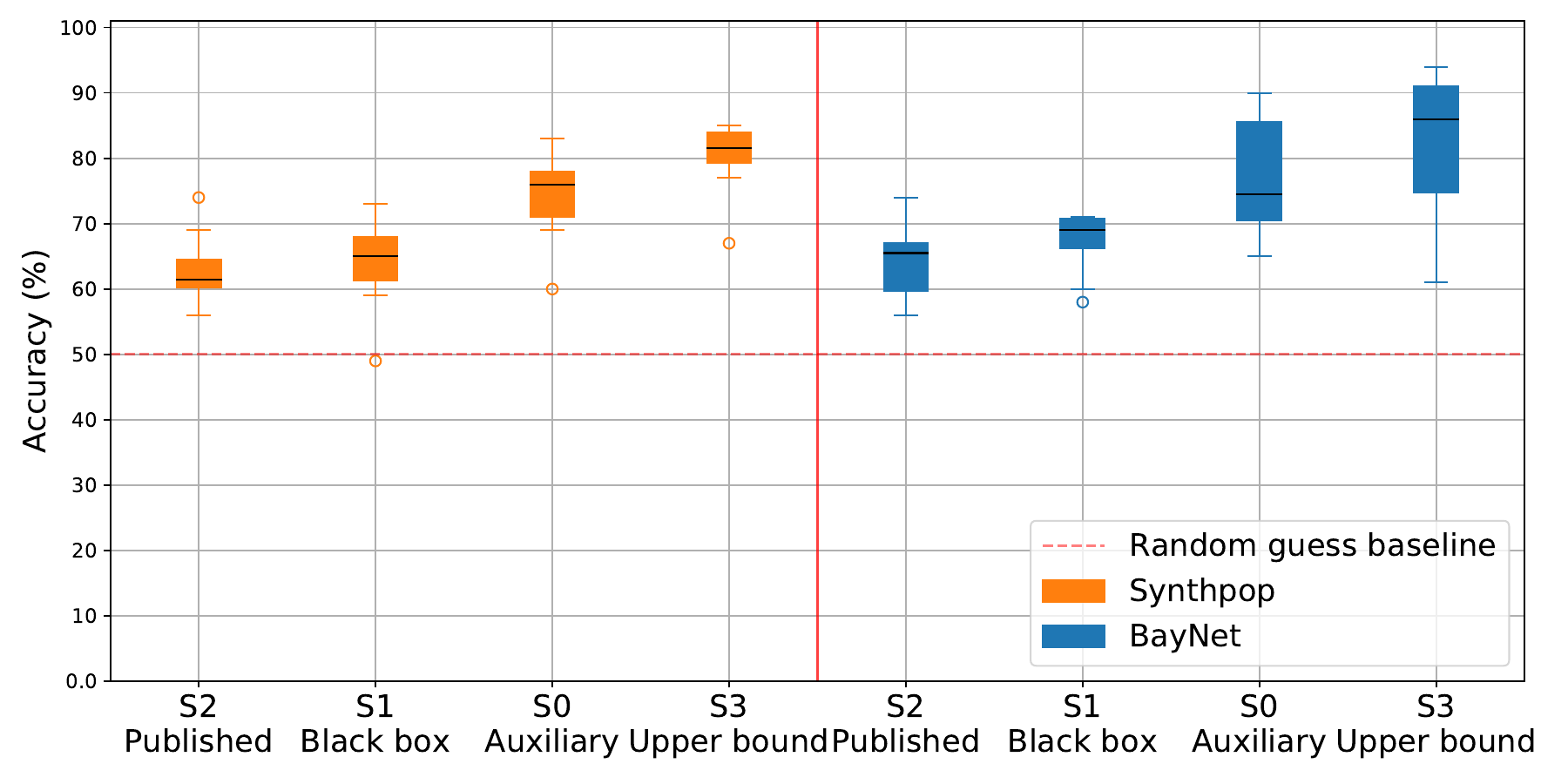}
}
    \caption{Comparison of MIA accuracy for the query based attack method across the 4 different scenarios (S0, S1, S2 and S3), for both generators Synthpop and BayNet. Figure (a) shows results for UK Census, while figure (b) displays results for Adult.} 
\label{fig:queries_figure}
\end{figure}

\subsection{Query based attack}
\label{sec:query_based_results}

We first use the state-of-the-art, query based attack method as introduced by Houssiau et al. \cite{houssiau2022tapas} to compare the MIA performance across different scenarios. 

We start by evaluating our weak attackers (S1) Black box and (S2) Published, where the attacker has only access to the target generator $\Phi_T$ or the released synthetic dataset respectively.

Figure \ref{fig:queries_figure} and table \ref{tab:all_results_queries} show that, across datasets and generators, an attacker in the (S1) Black box scenario achieves an average accuracy of 65.5\%. 
This is 15.5\% better than the random guess baseline of 50\%.
This shows that the traditional, strong assumption of having access to an auxiliary dataset can be removed while still successfully inferring membership. 

Next, we aim to make the attack as realistic as possible. To achieve that goal, we weaken the assumptions for the attacker to only have access to the published synthetic dataset ((S2) Published). Remarkably, we find that the MIA performance remains fairly constant when compared to the (S1) Black box scenario. Figure~\ref{fig:queries_figure} shows that across datasets and generators, we achieve an average accuracy of 62.8\%, which is only 2.7 p.p. lower than the (S1) Black box scenario. These results show that MIAs against synthetic data can still be successful, i.e. 12.8 p.p. better than the random baseline, when the released dataset is the only information available to the attacker. Given that releasing synthetic data instead of the original dataset is often the ultimate goal of generating synthetic data, we argue that scenario (S2) Published represents a minimal assumption that is almost always met in practice. Our results show that even in this realistic case,
records detected by the vulnerable record identification method of Meeus et al. \cite{meeus2023achilles} are at risk. 

\begin{table}[!ht]
  \centering
  \caption{MIA accuracy results (mean and standard deviation for 10 target records) across datasets and generators, for the query based attack.}
  \begin{tabular}{|l|c|c|c|c|c|}
    \hline
    \multicolumn{1}{|c|}{\multirow{2}{*}{\textbf{Scenario}}} & \multicolumn{2}{c|}{\textbf{UK census}}&  \multicolumn{2}{c|}{\textbf{Adult}} & \multirow{2}{*}{\textbf{Average}}\\
    &\multicolumn{1}{c}{Synthpop}&BayNet&\multicolumn{1}{c}{Synthpop}& BayNet &\\
    \hline
    S0: Auxiliary & $78.6 \pm 3.5$ \% & $78.4 \pm 3.4$ \% & $74.3 \pm 6.2$ \% & $77.0 \pm 8.6$ \% & $77.1 \pm 5.4$ \%\\
    \hline
    S1: Black-Box & $66.3 \pm 3.0$ \% & $64.6 \pm 5.3$ \% & $64.1 \pm 6.6$ \% & $67.2 \pm 4.4$ \% & $65.6 \pm 4.8$ \%\\
    \hline
    S2: Published & $61.9 \pm 3.3$ \% & $61.8 \pm 3.3$ \% & $63.1 \pm 4.9$ \% & $64.4 \pm 5.1$ \% & $62.8 \pm 4.2$ \%\\
    \hline
    S3: Upper Bound & $91.1 \pm 4.0$ \% & $89.3 \pm 5.0$ \% & $80.3 \pm 5.1$ \% & $82.5 \pm 1.1$ \% & $85.8 \pm 3.8$ \%\\
    \hline
  \end{tabular}
  \label{tab:all_results_queries}
\end{table}

We then compare the performance of our weak attacker (S1) Black-box to the traditional strong attacker (S0), where we assume access to an auxiliary dataset $D_{aux}$ of real records from the same distribution as the target dataset $D$.
Figure \ref{fig:queries_figure} shows that our (S1) attacker achieves an accuracy 11.6 p.p. lower compared to the baseline scenario (S0), on average across datasets and generators.
This is expected for two possible reasons.
First, the synthetic data might not be perfectly representative of the original distribution $\mathcal{D}$.
Thus, the training distribution of the meta-classifier from scenario (S1), consisting of features extracted from shadow generators trained on not perfectly representative data, might be quite different from the one on which it is evaluated,leading to worse performance.
Scenario (S0) does not suffer from this issue, since the meta-classifier is trained on features extracted from shadow generators trained on subsets of $D_{aux}$, which was itself sampled from the underlying distribution $\mathcal{D}$.
Second, there is the potential double counting issue, which we investigate next. 

Figure~\ref{fig:queries_figure} shows that (S3) Upper Bound achieves an MIA performance of 20.3 p.p. more than (S1) Black Box. These results suggest that the double issue issue might be significantly affecting the performance of the weaker attackers and that fixing this issue could, in the future, bridge the gap between our weak attackers and the (S3) Upper bound scenario.

Lastly, we find that on average, as reported in table \ref{tab:all_results_queries} across datasets and generators, this attacker achieves an accuracy of 85.8\%, which is 8.7 p.p. higher than in the (S0) Auxiliary scenario. This suggests that synthetic data is representative enough to construct shadow datasets for a successful MIA, and potentially more representative than an auxiliary dataset allowing to outperform scenario (S0). 

\subsection{Target attention attack}
\begin{figure}[!ht]
\centering

\subfigure[UK Census]{
\includegraphics[width=0.45\linewidth]{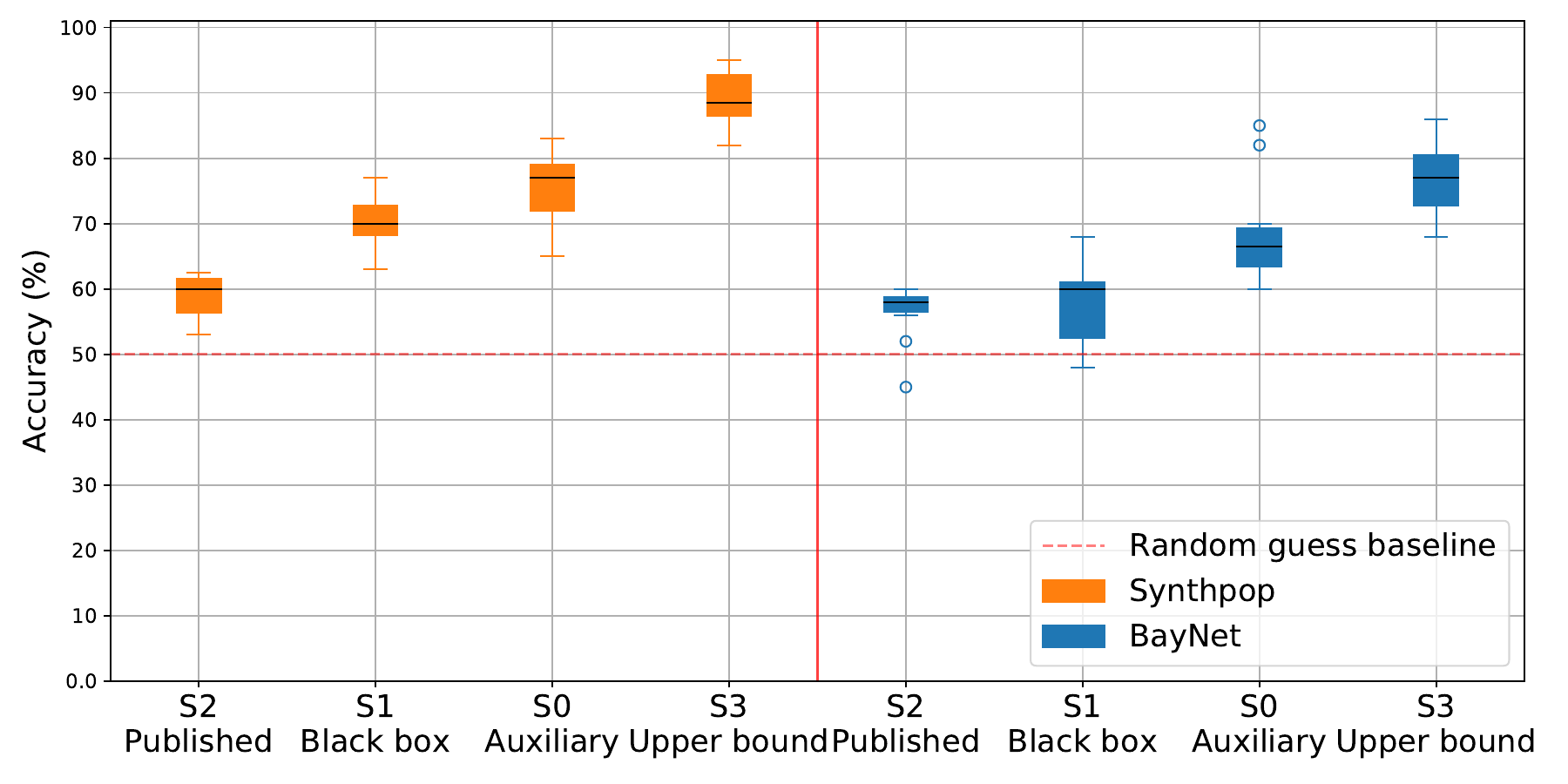}
}
\subfigure[Adult]{
\includegraphics[width=0.45\linewidth]{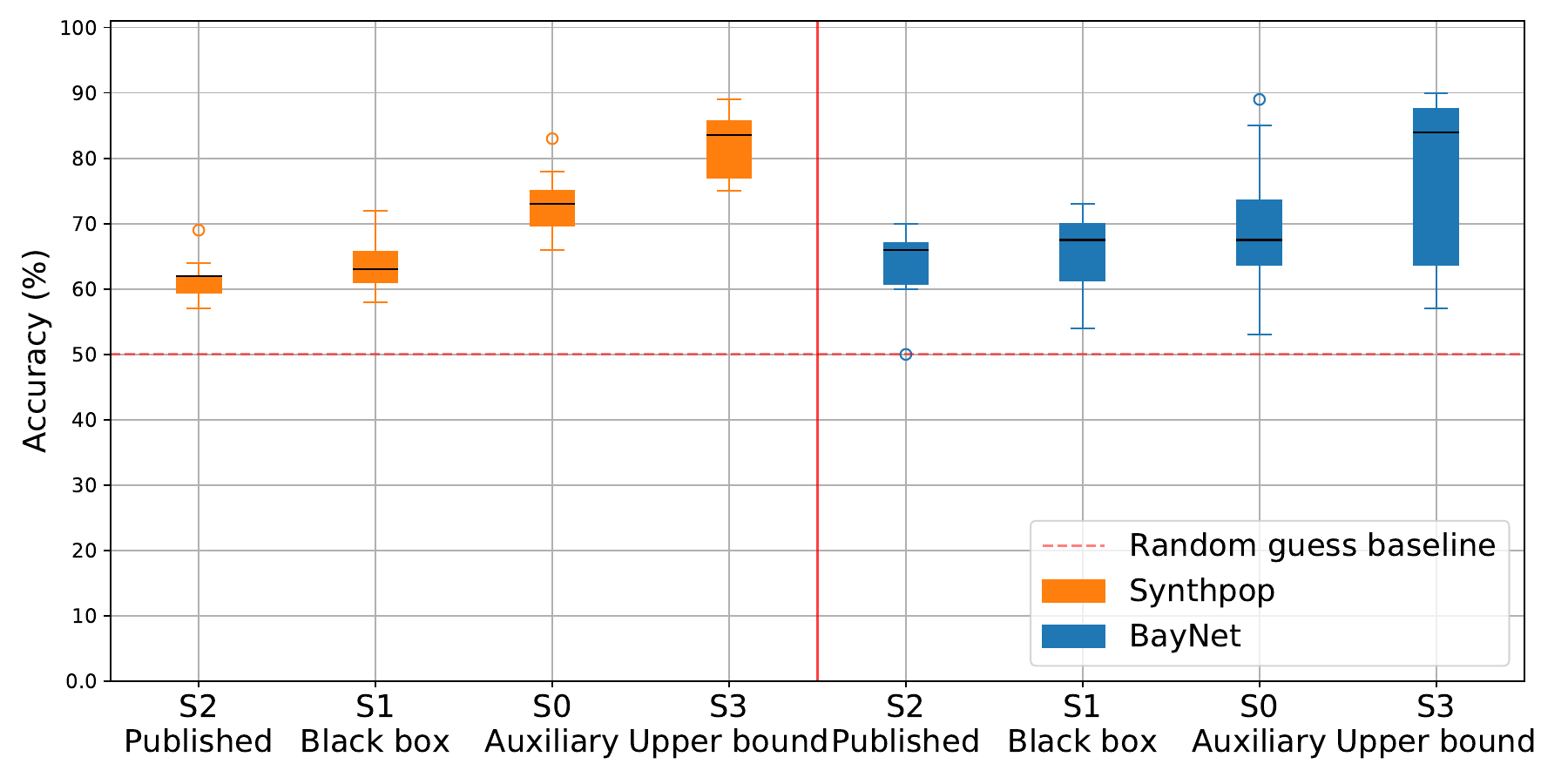}
}
    \caption{Comparison of MIA accuracy for the target attention attack method across the 4 different scenarios S0, S1, S2 and S3, for both generators Synthpop and BayNet. Figure (a) shows results for UK Census, while figure (b) displays results for Adult.} 
\label{fig:attention_figure}
\end{figure}

In this section, we evaluate if our results and conclusion are consistent across attack methods. To do this, we run the target attention attack method as proposed by Meeus et al. \cite{meeus2023achilles} using the same generator and datasets, with the same attack scenarios (S0-3). 
Figure \ref{fig:attention_figure} and table \ref{tab:all_results_attention} summarize the results. 

First, we find that the attacker from scenario (S1) is still successful using the target attention attack. 
Across datasets and generators, the average accuracy of the MIA lies at 63.3 \%, which is 13.3 p.p. better than the random baseline. This confirms that after removing access to the auxiliary dataset, records remain vulnerable against MIAs, even when using a distinct attack method.

Second, in scenario (S2), the MIA using the target attention method achieves 60.2\%, a drop of 3.1 p.p. compared to scenario (S1). These results show that the most realistic scenario, even across attack methods, can be considered as a realistic threat with a performance significantly above the random guess baseline.

Next, we find that the difference between the baseline scenario (S0) Auxiliary and scenario (S1) is on par with the results for the query-based attack. Across datasets and generators, the average accuracy drops by 8.5 p.p. while still achieving an average score of 63.3\%.  

Finally, in scenario (S3), we confirm our findings that the double counting issue is the main reason affecting the performance of the weaker attackers, also when using the target attention method. The MIAs achieve an average of 81.2\% accuracy, which is 9.4 p.p. higher than (S0) and 17.9 p.p. higher than (S1). 

The fact that our findings are consistent across two very distinct attack methods suggests that even when new attack methods are developed, MIAs against synthetic data using only synthetic data will be successful. 

\begin{table}[!ht]
  \centering
  \caption{MIA accuracy results (mean and standard deviation for 10 target records) across datasets and generators, for the target attention attack.}
  \begin{tabular}{|l|c|c|c|c|c|}
    \hline
    \multicolumn{1}{|c|}{\multirow{2}{*}{\textbf{Scenario}}} & \multicolumn{2}{c|}{\textbf{UK census}}&  \multicolumn{2}{c|}{\textbf{Adult}} & \multirow{2}{*}{\textbf{Average}}\\
    &\multicolumn{1}{c}{Synthpop}&BayNet&\multicolumn{1}{c}{Synthpop}& BayNet & \\
    \hline
    S0: Auxiliary & $75.4 \pm 5.4$ \% & $68.7 \pm 7.9$ \% & $73.2 \pm 4.7$ \% & $69.7 \pm 10.3$ \% & $71.8 \pm 7.1$ \%\\
    \hline
    S1: Black-Box & $61.5 \pm 3.3$ \% & $62.1 \pm 6.3$ \% & $64.1 \pm 4.3$ \% & $65.5 \pm 6.2$ \% & $63.3 \pm 5.0$ \%\\
    \hline
    S2: Published & $58.9 \pm 3.0$ \% & $56.4 \pm 4.3$ \% & $61.5 \pm 3.3$ \% & $63.8 \pm 5.6$ \% & $60.2 \pm 4.1$ \%\\
    \hline
    S3: Upper Bound & $88.9 \pm 4.4$ \% & $76.8 \pm 5.2$ \% & $82.0 \pm 4.9$ \% & $77.2 \pm 13.0$ \% & $81.2 \pm 6.9$ \%\\
    \hline
  \end{tabular}
  \label{tab:all_results_attention}
\end{table}

\subsection{Robustness analysis for number of synthetic records $m$}

In scenario (S1) Black Box, we assume the attacker to have black box access to the target generator, i.e. the attacker can query the generator for synthetic records an arbitrary number of times. In our experiments we set the number of synthetic records $m$ to $20,000$. 

\begin{figure}[!ht]
\centering
\subfigure[]{
\includegraphics[width=0.3\linewidth]{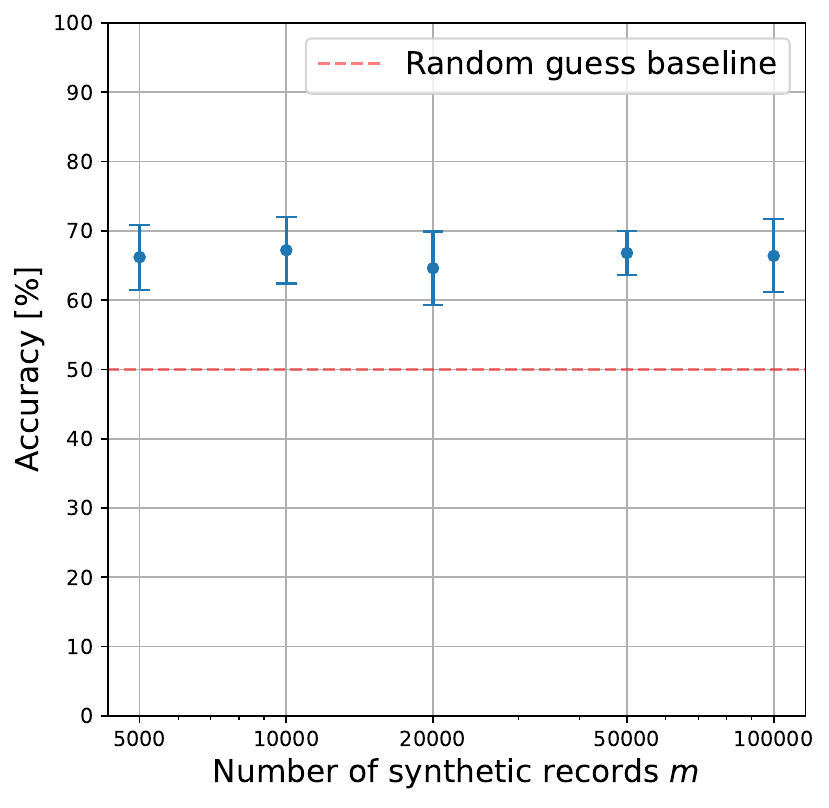}}
\subfigure[]{
\includegraphics[width=0.3\linewidth]{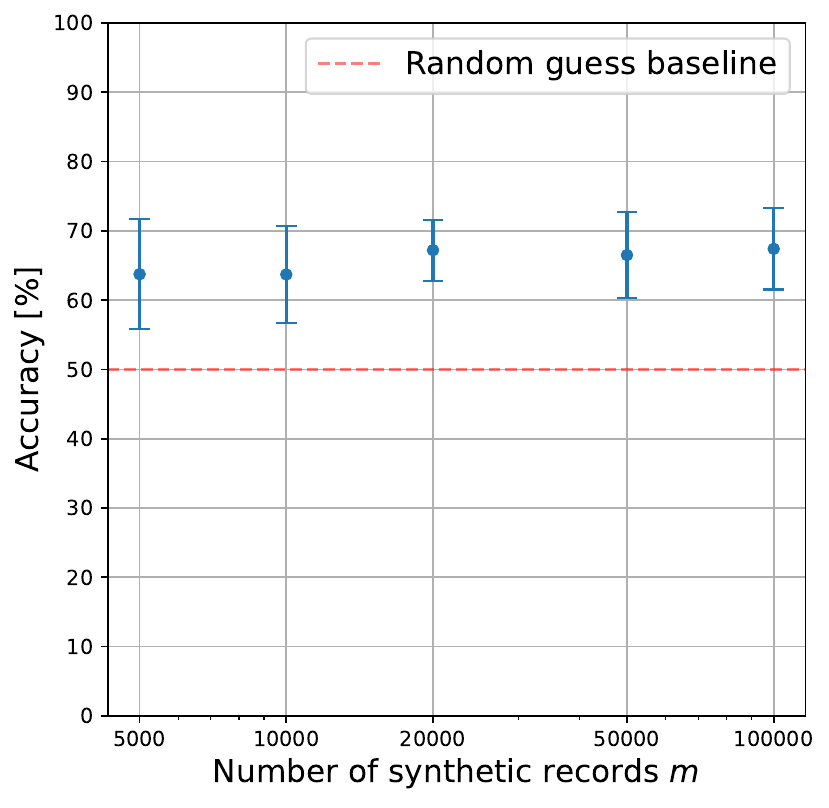}
}
    \caption{Mean and standard deviation of MIA accuracy for scenario (S1) Black-Box for varying number $m$ synthetic records available to the attacker. Results for BayNet and the query-based attack using (a) UK Census (b) Adult.} 
\label{fig:varying_m}
\end{figure}

We now evaluate the effect of the value of $m$ on the attack performance. Across the two datasets, for the BayNet generator, Figure \ref{fig:varying_m} shows how the MIA performance for scenario (S1) varies for increasing $m$. 

Across datasets, the MIA accuracy remains fairly robust for varying number of synthetic records made available to the attacker. For $m$ varying from $5,000$ to $100,000$, the mean MIA accuracy does not change significantly. First, this shows that $m=20,000$, as used in our experimental setup, is a good approximation for black box access to the generator. Further, along with the MIA results for scenario (S2) Published, this confirms that releasing a number of synthetic records $m$ larger or equal to the number of original records, allows the attacker to build meaningful MIAs. 

%% file: sections/future_work.tex
\subsection{Impact of releasing less synthetic records}
Intuitively, for a training dataset of fixed size, the more synthetic records we generate, the more information the synthetic dataset might start leaking. 

In scenarios (S0) and (S2), the attacker only has access to a limited number of synthetic records $m=|\mathcal{D}|$. As synthetic data is often used to replace the original dataset, we argue that it is reasonable in practice to generate the same amount of synthetic records as the number of training records.

However, we hypothesize that releasing fewer synthetic records for a fixed size of the training dataset, namely $m < |\mathcal{D}|$, could reduce the accuracy of our attack. Of course, releasing less synthetic records typically comes at a cost in utility. We leave the evaluation of this potential trade-off between $m$ and the accuracy of our attack on the released synthetic data for future work.

\subsection{Differentially private synthetic generation methods}

As main contribution in this work, we show that it is possible to attack a synthetic data generator based only on the generated synthetic data. 

We leave for future work to confirm whether these effects translate to synthetic data generators with formal privacy guarantees, such as differentially private generators proposed in the literature \cite{zhang2017,jordon2019pate}. Previous work has shown that Differential Privacy (DP) comes at a cost in utility \cite{stadler2022synthetic,annamalai2023linear} and that the MIA accuracy drops for decreasing value of the privacy budget $\epsilon$ \cite{meeus2023achilles}. We expect that, while exhibiting similar trends, our findings would translate to DP generators. 

\subsection{Bridging the gap with the upper bound}

Our results show that scenario (S3) achieves a significant MIA accuracy, namely higher than scenarios (S1) and (S2), and even higher than scenario (S0). We leave for future work to address the double counting issue we identified in practice, to bridge the gap between scenarios (S1, S2) and the upper bound scenario (S3).

Potentially, an attacker could remove the synthetic records close to the target record, prior to using the synthetic data to construct the shadow models. This could reduce the impact of the double counting issue, but might also introduce bias into the shadow datasets. 
Additionally, note that in scenario (S1) we currently train the meta classifier using shadow datasets randomly sampled from $m=20000$ synthetic records, to then infer a prediction on a random subset of $n_{synthetic} = 1000$ synthetic records. An attacker could for instance infer the prediction on multiple subsets of the $m$ synthetic records to potentially make a more optimal, ensemble prediction.

%% file: sections/conclusion.tex
Sharing data plays a pivotal role in research and innovation. Increasingly, synthetic data has been proposed to share privacy-preserving tabular data, by synthesizing records that are not directly linkable to real records, while retaining data utility. 

Membership Inference Attacks (MIAs) are the standard to audit the privacy preservation of synthetic data, and recent work has shown that these attacks can successfully infer the membership of certain records in the private dataset. State-of-the-art MIAs rely on shadow modeling, which traditionally assumes an attacker to have access to an auxiliary dataset.

 First, this auxiliary data assumption is hard to meet in practice. 
 Second, GDPR Recital 26 \cite{eu-gdpr} states that, to legally meet anonymization standards, all means reasonably likely for an attacker to possess should be considered. 

We here proposed a more realistic attack by removing the auxiliary data assumption. Across two real world datasets and two synthetic data generators, we find that MIAs are still successful when only using synthetic data. 

Specifically, we find that on average, an attacker with black box access to the generator achieves 65.5\% accuracy, while an attacker with only access to the released synthetic dataset attains an accuracy of 62.8\%. The latter result is particularly significant as it demonstrates that an attacker can extract sensitive information from released synthetic data without any additional information.

Moreover, we identify a double counting issue and establish an upper bound for MIA accuracy against synthetic data when only synthetic data is available. Using current state-of-the-art attacks, this upper bound stands at 85.8\%, which is, remarkably, higher than traditional attacks using auxiliary data. This finding highlights the potential for future researchers to bridge the existing gap of MIA performance between realistic scenarios and the upper bound. 

Our results provide compelling evidence that MIAs against synthetic data pose a realistic threat in practice. We hope this helps researchers and practitioners to better evaluate risks associated with releasing synthetic data, while encouraging the development of methods to address these concerns.